\documentclass[useAMS,usenatbib]{mn2e}
\usepackage{amsmath,amssymb,epsfig,natbib,bm,psfrag,ifthen}
\usepackage[hypertex]{hyperref}
\begin{document}



\def\eprinttmp@#1arXiv:#2 [#3]#4@{
\ifthenelse{\equal{#3}{x}}{\href{http://arxiv.org/abs/#1}{#1}}{\href{http://arxiv.org/abs/#2}{arXiv:#2} [#3]}}

\providecommand{\eprint}[1]{\eprinttmp@#1arXiv: [x]@}
\newcommand{\adsurl}[1]{\href{#1}{ADS}}
\providecommand{\bibinfo}[2]{\ifthenelse{\equal{#1}{isbn}}{%
\href{http://cosmologist.info/ISBN/#2}{#2}}{#2}}
\providecommand{\ISBN}[1]{ISBN: \href{http://cosmologist.info/ISBN/#1}{#1}}

\newcommand{\ud}{{\rm d}}

\renewcommand{\tilde}{\widetilde}

\newcommand\Tr{{\rm Tr}}

\newcommand{\HALOFIT}{\textsc{halofit}}
\newcommand{\Mpc}{\text{Mpc}}
\newcommand{\half}{{\textstyle \frac{1}{2}}}
\newcommand{\third}{{\textstyle \frac{1}{3}}}
\newcommand{\numfrac}[2]{{\textstyle \frac{#1}{#2}}}
\newcommand{\ra}{\rangle}
\renewcommand{\la}{\langle}
\renewcommand{\d}{\text{d}}
\newcommand{\grad}{\mbox{\boldmath$\nabla$}}

\newcommand{\begm}{\begin{pmatrix}}
\newcommand{\enm}{\end{pmatrix}}

\newcommand{\threej}[6]{{\begm #1 & #2 & #3 \\ #4 & #5 & #6 \enm}}
\newcommand{\fsky}{f_{\text{sky}}}

\newcommand{\cla}{\mathcal{A}}
\newcommand{\clb}{\mathcal{B}}
\newcommand{\clc}{\mathcal{C}}
\newcommand{\cle}{\mathcal{E}}
\newcommand{\clf}{\mathcal{F}}
\newcommand{\clg}{\mathcal{G}}
\newcommand{\clh}{\mathcal{H}}
\newcommand{\cli}{\mathcal{I}}
\newcommand{\clj}{\mathcal{J}}
\newcommand{\clk}{\mathcal{K}}
\newcommand{\cll}{\mathcal{L}}
\newcommand{\clm}{\mathcal{M}}
\newcommand{\cln}{\mathcal{N}}
\newcommand{\clo}{\mathcal{O}}
\newcommand{\clp}{\mathcal{P}}
\newcommand{\clq}{\mathcal{Q}}
\newcommand{\clr}{\mathcal{R}}
\newcommand{\cls}{\mathcal{S}}
\newcommand{\clt}{\mathcal{T}}
\newcommand{\clu}{\mathcal{U}}
\newcommand{\clv}{\mathcal{V}}
\newcommand{\clw}{\mathcal{W}}
\newcommand{\clx}{\mathcal{X}}
\newcommand{\cly}{\mathcal{Y}}
\newcommand{\clz}{\mathcal{Z}}
\newcommand{\CMBFAST}{\textsc{cmbfast}}
\newcommand{\CAMB}{\textsc{camb}}
\newcommand{\COSMOMC}{\textsc{CosmoMC}}
\newcommand{\Healpix}{\textsc{healpix}}
\newcommand{\IGLOO}{\textsc{igloo}}
\newcommand{\GLESP}{\textsc{glesp}}

\newcommand{\Omtot}{\Omega_{\mathrm{tot}}}
\newcommand{\Omb}{\Omega_{\mathrm{b}}}
\newcommand{\Omc}{\Omega_{\mathrm{c}}}
\newcommand{\Omm}{\Omega_{\mathrm{m}}}
\newcommand{\omb}{\omega_{\mathrm{b}}}
\newcommand{\omc}{\omega_{\mathrm{c}}}
\newcommand{\omm}{\omega_{\mathrm{m}}}
\newcommand{\Omdm}{\Omega_{\mathrm{DM}}}
\newcommand{\Omnu}{\Omega_{\nu}}

\newcommand{\Oml}{\Omega_\Lambda}
\newcommand{\OmK}{\Omega_K}

\newcommand{\Hunit}{~\text{km}~\text{s}^{-1} \Mpc^{-1}}
\newcommand{\Gyr}{{\rm Gyr}}

\newcommand{\nrun}{n_{\text{run}}}

\newcommand{\lmax}{l_{\text{max}}}

\newcommand{\zre}{z_{\text{re}}}
\newcommand{\mpl}{m_{\text{Pl}}}

\newcommand{\vphi}{\mathbf{\psi}}
\newcommand{\vv}{\mathbf{v}}
\newcommand{\vI}{\mathbf{I}}
\newcommand{\vm}{\mathbf{m}}

\newcommand{\vd}{\mathbf{d}}
\newcommand{\vC}{\mathbf{C}}
\newcommand{\vT}{\mathbf{T}}
\newcommand{\vX}{\mathbf{X}}
\newcommand{\vn}{\hat{\mathbf{n}}}
\newcommand{\vy}{\mathbf{y}}
\newcommand{\vk}{\mathbf{k}}

\newcommand{\mN}{\bm{N}}
\newcommand{\eV}{\,\text{eV}}
\newcommand{\vtheta}{\bm{\theta}}
\newcommand{\tT}{\tilde{T}}
\newcommand{\tE}{\tilde{E}}
\newcommand{\tB}{\tilde{B}}

\newcommand{\mCh}{\hat{\bm{C}}}
\newcommand{\Ch}{\hat{C}}

\newcommand{\Bt}{\tilde{B}}
\newcommand{\Et}{\tilde{E}}
\renewcommand{\bld}[1]{\mathrm{#1}}
\newcommand{\mLambda}{\bm{\Lambda}}
\newcommand{\mA}{\bm{A}}
\newcommand{\mpsi}{\bm{\Psi}}

\newcommand{\mC}{\bm{C}}
\newcommand{\mQ}{\bm{Q}}
\newcommand{\mU}{\bm{U}}
\newcommand{\mX}{\bm{X}}
\newcommand{\mV}{\bm{V}}
\newcommand{\mP}{\bm{P}}
\newcommand{\mR}{\bm{R}}
\newcommand{\mW}{\bm{W}}
\newcommand{\mD}{\bm{D}}
\newcommand{\mI}{\bm{I}}
\newcommand{\mH}{\bm{H}}
\newcommand{\mM}{\bm{M}}
\newcommand{\mS}{\bm{S}}
\newcommand{\mzero}{\bm{0}}
\newcommand{\mL}{\bm{L}}

\renewcommand{\btheta}{\bm{\theta}}
\renewcommand{\bphi}{\bm{\psi}}

\newcommand{\vb}{\mathbf{b}}
\newcommand{\vA}{\mathbf{A}}
\newcommand{\vG}{\mathbf{G}}

\newcommand{\vAt}{\tilde{\mathbf{A}}}
\newcommand{\ve}{\mathbf{e}}
\newcommand{\vE}{\mathbf{E}}
\newcommand{\vB}{\mathbf{B}}
\newcommand{\vEt}{\tilde{\mathbf{E}}}
\newcommand{\vBt}{\tilde{\mathbf{B}}}
\newcommand{\vEw}{\mathbf{E}_W}
\newcommand{\vBw}{\mathbf{B}_W}
\newcommand{\vx}{\mathbf{x}}
\newcommand{\vXt}{\tilde{\vX}}
\newcommand{\vXb}{\bar{\vX}}
\newcommand{\vTb}{\bar{\vT}}
\newcommand{\vTt}{\tilde{\vT}}
\newcommand{\vY}{\mathbf{Y}}
\newcommand{\vBwr}{{\vBw^{(R)}}}
\newcommand{\RW}{{W^{(R)}}}

\newcommand{\mUt}{\tilde{\mU}}
\newcommand{\mVt}{\tilde{\mV}}
\newcommand{\mDt}{\tilde{\mD}}

\newcommand{\Rot}{\begm \mzero &\mI \\ -\mI & \mzero \enm}
\newcommand{\Pt}{\begm \vEt \\ \vBt \enm}

\newcommand{\edth}{\,\eth\,}
\renewcommand{\beth}{\,\overline{\eth}\,}

\newcommand{\fpsf}{f_{\text{psf}}}

\newcommand{\sE}{{}_{|s|}E}
\newcommand{\sB}{{}_{|s|}B}
\newcommand{\sElm}{\sE_{lm}}
\newcommand{\sBlm}{\sB_{lm}}

\newcommand{\alt}{\lesssim}
\newcommand{\agt}{\gtrsim}

\title{Galaxy shear estimation from stacked images}

\author[A.M.~Lewis]{Antony Lewis$^{1}$\thanks{Contact details at \url{http://cosmologist.info}}\\$^1$Institute of
Astronomy, Madingley Road, Cambridge CB3 0HA, UK.}

\maketitle

\begin{abstract}

Statistics of the weak lensing of galaxies can be used to constrain cosmology if the galaxy shear can be
estimated accurately. In general this requires accurate modelling of unlensed galaxy shapes and
the point spread function (PSF). I discuss suboptimal but potentially robust methods for estimating galaxy shear by stacking images such that the stacked image distribution is closely Gaussian by the central limit theorem. The shear can then be determined by radial fitting, requiring only an accurate model of the PSF rather than also needing to model each galaxy accurately. When noise is significant asymmetric errors in the centroid must be corrected, but the method may ultimately be able to give accurate un-biased results when there is a high galaxy density with constant shear. It provides a useful baseline for more optimal methods, and a test-case for estimating biases,
though the method is not directly applicable to realistic data. I test stacking methods on the simple toy simulations with constant PSF and shear provided by the GREAT08 project, on which most other existing methods perform significantly more poorly, and briefly discuss generalizations to more realistic cases.
In the appendix I discuss a simple analytic galaxy population model where stacking gives optimal errors in a perfect ideal case.
\end{abstract}

\section{Introduction}

Gravitational lensing of light from distant galaxies causes the shape of the galaxies to be distorted in a way that depends on the transverse gradients of the gravitational potential along the line of sight~\citep{Bartelmann:1999yn}. If the distortion can be measured accurately, it gives a constraint on the lensing potentials, and hence with large enough number of samples on the geometry and distribution of perturbations in the universe. Since the galaxy shapes vary greatly, this can only be done by analysing a very large number of galaxies, with galaxies that are sufficiently well separated that their intrinsic shape correlations can be modelled out or is small. The galaxies can then be assumed to be independent, so that any shape correlation is due entirely to lensing. The task is to find a way to estimate the lensing distortion, which can then be used to extract statistical results from an ensemble of galaxy images.

At leading order the main observable distortion is that of galaxy shear. As discussed further below, if we could observe the galaxies directly, fitting any sheared profile to each galaxy will give an unbiased estimator of this shear. The problem is however much more complicated, because in practice we can only measure the shape after convolution with the point spread function (PSF) of the instrument (e.g. due to atmospheric fluctuations and instrumental imperfections), and image pixelization. The levels of shear that are expected --- a few percent --- are comparable to those of typical PSFs, so the PSF must be modelled very accurately in order to isolate the cosmological signal. Since the PSF breaks the symmetries of the problem, in general this requires accurate modelling of both the unlensed galaxy shapes and the PSF. Finding methods of doing this that work to the required level of precision is an active area of current research. At the moment it unclear whether it is even possible to get useful high-precision shear constraints in the presence of realistic ground-observation PSFs, or whether in fact there are unavoidable degeneracies with galaxy shapes and PSF modelling uncertainties. The correct statistical error on the shear measurement could also be too large for the number of observable galaxies to produce precision constraints.

In this paper I re-visit an old sub-optimal method of~\citet{Kuijken:1999jz}: stacking galaxy images. If the intrinsic galaxy shapes are uncorrelated, a stacked unlensed image should have circular symmetry. Since convolution is a linear operation, the observed stacked image should then be a PSF-convolved sheared version of a circularly symmetric average galaxy. If the PSF is known, the only modelling uncertainties are then in the averaged galaxy profile, which should be well determined by the data. Furthermore, under fairly general conditions a sum of independent samples should have a close-to-Gaussian distribution by the central limit theorem, so the statistics of the stacked image is known without needing to know anything about the distribution of individual galaxy shapes. Fitting a radial profile and shear to the data with a Gaussian error model gives an estimate of the shear that should be very independent of the actual distribution of galaxy shapes. The method therefore provides a useful baseline for comparing future more optimal methods that incorporate accurate modelling of individual galaxies.

In practice of course things are not so simple. To start with the shear and PSF are not expected to be constant, so any stacked galaxy image has to be interpreted with care. In addition, in the presence of noise, the process of stacking can itself produce biases since we cannot determine the centroid accurately: any shear- or PSF-correlated misalignments in the stacking procedure will introduce biases.

Given the complexity of the general problem, the lensing community has helpfully boiled the issues down into a series of much simpler problems~\citep{Heymans:2005rv,Massey:2006ha,Bridle:2008iv}. Although existing methods perform adequately for current and near-future data, even in these highly simplified cases they are known to be inadequate for future surveys. I therefore focus on these simplified problems to try to isolate the important issues, in particular I shall assume the PSF is well measured from many low-noise star images and that shear is constant. If no methods works accurately even on this very simple toy problem, then that is clearly sufficient to show that ground-based weak lensing surveys with similar PSFs will be of no use for precision cosmology (i.e. future cosmological parameter constraints at the percent level or better). On the other hand if sufficiently accurate methods can be developed, the next task will be to make them applicable to more realistic situations where the PSF is likely to vary significantly and the shear has a realistic spatial correlation function. Space-based observations typically have rather different PSFs and would require a separate study.

I start by reviewing the case of shape estimation when there is no PSF, and then briefly explain why introducing a PSF qualitatively increases the complexity of the problem. I then move on to show that stacking can work well with low-noise simulations, and discuss various issues to do with pixel-scale stacking, centroid errors and non-constant PSFs. I test stacking methods on the GREAT08\footnote{\url{http://www.great08challenge.info/}} simulations~\citep{Bridle:2008iv} and show that it performs well compared to other existing methods, most of which involve modelling unlensed galaxy shape distributions with something that is known to be incorrect.  Unlike other existing methods the stacking method is not directly applicable to more realistic data, but may be useful to motivate more general approaches.

\section{Fitting galaxies and shear}

\subsection{Shear fitting with no PSF}

At lowest order in the gravitational potentials weak lensing causes position $\vx_u$ on the unlensed image to be related to the corresponding position $\vx_l$ on the lensed image by
\begin{equation}
\vx_u = \mS\vx_l \equiv \begm 1-g_1 && -g_2 \\ -g_2 && 1+g_1\enm \vx_l,
\end{equation}
where the components of the shear matrix $g_1$ and $g_2$ are the reduced shear in some coordinate system. For the purposes of this paper we can neglect the uniform convergence which is degenerate with the galaxy size and assume $g_1, g_2$ are constant across each galaxy image. For a thorough introduction to weak lensing see~\citet{Bartelmann:1999yn,Lewis:2006fu}.

Consider fitting a model $m(\mS_m\vx,\theta)$  to an unlensed perfect galaxy image $I_u(\vx)$, with model parameters $\theta$ and shear matrix $\mS_m$.
 For example a simple least-squares fit would solve
\begin{equation}
\frac{\partial}{\partial\theta}\int d^2\vx |I_u(\vx) - m(\mS_m\vx,\theta)|^2 = 0 \nonumber
\end{equation}
\begin{equation}\frac{\partial}{\partial\mS_m}\int d^2\vx |I_u(\vx) - m(\mS_m\vx,\theta)|^2 = 0.
\end{equation}
Assuming there is a unique solution $\mS_m = \mS_0, \theta = \hat{\theta}$, the lensed image $I_l(\vx) = I_u(\mS\vx)$ would then be fit by $m(\hat{\mS}\vx,\hat{\theta})$ where $\hat{\mS} = \mS_0\mS$. The best-fit unlensed shear matrix $\mS_0$ is determined by the particular galaxy and model. The key assumption in galaxy weak lensing is that the galaxy shapes are statistically isotropic, in other words that versions of each unlensed galaxy rotated by different angles (or flipped) are equally likely. So $\mR^T\mS_0 \mR$ is just as likely as $\mS_0$ for some rotation matrix $\mR$. Taking $\mR$ to be a rotation by $90^\circ$, on average over many galaxy orientations
we have, $\la \mS_0\ra = \half\la \mS_0 + \mR^T\mS_0\mR\ra =  \mI$, and hence $\la \hat{\mS}\ra  = \mS$: the shear matrix estimator is unbiased. Note that this is entirely independent of how well $m(\mS_m\vx,\theta)$ actually fits the unlensed galaxy, so in the idealized case we could fit any model we like to galaxy shapes and still on average get the correct answer. This will remain true for best-fits to more general log-likelihoods of the form
\begin{multline}
\chi^2 = \int d^2\vx \left(I_l(\vx) - m({\mS}_m\vx, \theta)\right)^T [\mN(I_l(\vx),{\mS}_m\vx)]^{-1}\\
\times \left(I_l(\vx) - m({\mS}_m \vx, \theta)\right),
\end{multline}
i.e. where the noise depends only on the lensed galaxy intensity or follows the alignment of the galaxy model. Similarly for generalizations with correlated noise.

\subsection{Shear fitting with a PSF}

Unfortunately we cannot observed lensed galaxies directly, but only after convolution with an instrumental point spread function and pixelization. Pixelization can be though of as an additional contribution to the PSF, typically a convolution with a square window function, followed by sampling at the pixel centres. I shall discuss the PSF in this generalized sense, so that the observational data consists of a set of regularly-spaced samples of a PSF and pixel-convolved galaxy image.
The noise-free observed value at position $\vx$ on the image plane is then
\begin{equation}
I_o(\vx) = \int d^2 \vy P(\vx-\vy) I_l(\vy) ,
\end{equation}
where $P(\vx)$ is the total PSF, or simply $I_o = P\star I_l$. If we know the PSF function, we can fit a PSF-convolved galaxy model to the data; for example a least-squares solution would minimize
\begin{eqnarray}
\chi^2 &=& \int d^2\vx  \biggl(I_o(\vx) - \int d^2\vy P(\vx-\vy) m(\mS_m\vy,\theta)\biggr)^2 \nonumber\\
&=&\int\! d^2\vx  \biggl[\int \d^2\vy P(\vx-\vy)  [I_u(\mS\vy) -m(\mS_m\vy,\theta)]\biggr]^2.
\end{eqnarray}
If $\mS_m=\mS_0, \theta=\hat{\theta}$ is the best fit when there is no lensing, due to the position dependence of the PSF it is no longer in general the case that $\mS_m = \mS_0\mS, \theta=\hat{\theta}$ is the best fit to the lensed image. Hence unlike in the case with no PSF, there is no longer any guarantee that fitting is giving an unbiased estimate of the shear. The only general exception is if the model fits the galaxy exactly, so the best fit has $I_u(\mS\vy) = m(\mS_m\vy,\theta)$, in which case fitting is giving the right answer on average independent of the known PSF.

Extracting unbiased shear constraints by model fitting in the presence of a PSF therefore in general requires modelling the galaxies accurately. This poses several significant problems. A large galaxy lensing survey will have most of its galaxies near the edge of its resolution, therefore there is typically only limited high-quality data to constrain the properties of the bulk of the galaxies in the selection function. A general Bayesian model could use information from some well-resolved galaxies, and fit a general model for uncertainties, but given the large variation in galaxy alignment with respect to the line of sight, and wide variations in the intrinsic shapes, a general model is likely to involve a large number of parameters and require many images to constrain well. The galaxy model can also be constrained to some extent using all of the observed galaxies. But if the parameters can not all be well constrained by the data, it may be essential that the priors accurately represent the galaxy distributions in order to get unbiased answers. In addition any model with large numbers of parameters per galaxy is likely to become numerically time consuming. For an excellent discussion of many related issues see~\citet{Bernstein:2001nz,Hirata:2003cv}, and a summary of other existing methods in~\citet{Bridle:2008iv}. For promising recent results on Bayesian model fitting see~\citet{Miller:2007an,Kitching:2008pd}, however the galaxy model used in this method is still unrealistic and results, though significantly better than many other methods, are still not good enough for high-precision cosmology (see Sec.~\ref{sims}).

\section{Shear fitting stacked galaxies}

General galaxy fitting should provide the best constraints on the shear. However given the problems outlined above, and given potential difficulties in knowing whether the modelling is accurate enough, it would be useful to have a simple less-optimal but more robust shear-estimation method that is more directly independent of the details of the galaxy distribution. In simple test cases this would be a useful cross-check, and provide a baseline for the levels of residual noise that better methods should be able to beat.

The method I shall focus on simply stacks the galaxies, and then fits a sheared average galaxy model to the stacked image, following~\citet{Kuijken:1999jz}. If the PSF is known, this should give unbiased results conditional only on being able to stack in an unbiased way, and being able to model the radial profile of the averaged unlensed galaxy accurately. A 1-dimensional radial model is clearly much easier to fit that a full 2D galaxy shape distribution, and since the average galaxy is expected to have a smooth radial profile only a modest number of parameters should be required. These parameters are likely to be well constrained with a reasonable number of galaxies (and hence the results fairly independent of the priors).

In order to stack galaxies images, we need to be able to define a rule for the relative galaxy alignment, e.g. by defining a centroid in each image and then stacking the images so that their centroids are aligned. Assuming this can be done, we then have an observed stack of $N$ galaxy images
\begin{equation}
\hat{\bar{I}}_o(\vx) \equiv \frac{1}{N}\sum_{i=1}^N \beta_i I_{o,i}(\vx) =
\frac{1}{N}\sum_{i=1}^N [\beta_i P_i \star I_{l,i}](\vx),
\end{equation}
where $\beta_i$ are some weights and $P_i$ the PSF on galaxy $i$. Assuming the PSF is independent of the galaxy shape the expected value of the stacked image is
\begin{equation}
\la \hat{\bar{I}}_o\ra = \frac{1}{N}\sum_{i=1}^N [P_i \star \la\beta_i I_{l,i}\ra]
 =  \left(\frac{1}{N}\sum_{i=1}^N P_i\right) \star \bar{I}_{l,\beta} = \bar{P}\star  \bar{I}_{l,\beta}  ,
\end{equation}
where $\bar{I}_{l,\beta}$ is the average of a weighted galaxy. By symmetry, taking $\vx$ to have origin at the centroid, in the unlensed case
 $\bar{I}_{u,\beta}(\vx) =\bar{I}_{u,\beta}(|\vx|)$.  The average PSF $\bar{P}$ --- including pixelization --- is precisely what is observed from a large statistically equivalent set of star images (assuming stars are point sources and have the same PSF as the galaxies).

 Assuming the weights are independent of the shear and the shear is constant, the expectation of the stacked image is a sheared circularly-symmetric averaged galaxy, convolved with an average PSF. We can therefore proceed to fit a model to the observed stacked image, and if the radial profile can be fit accurately the method should be unbiased. In the appendix I discuss a simple analytic galaxy population model in which, for the ideal noise-free case, stacking with an appropriate weighting is in fact optimal.

 In the presence of noise, and with finite $N$ so that there is dispersion about the expectation value, we need an error model. One benefit of using stacked images is that this is well defined:
 assuming each galaxy is independent, the distribution of $\hat{\bar{I}}_o$, a sum of many independent galaxy samples, should be nearly Gaussian by the central limit theorem.

In fact we can apply any linear function to $\hat{\bar{I}}_o$, and the distribution will still be Gaussian with expectation given by the equivalent linear function of the averaged convolved galaxy. This can be useful for data compression, e.g. to re-pixelize, or expand in moments, etc, anything that's likely to encapsulate most of the useful information in fewer numbers. If the stacked image is generated at much higher pixel sampling than the original image, there will be a large number of pixel values and hence a huge number of galaxies required for the covariance estimate to be accurate. Also since the noise is correlated on the scale of the original pixel size, the covariance would be singular, so applying some linear re-pixelization or other linear compression matrix $\mM$ can be useful. Writing the set of sampled $\vx$ values as a vector, for a data vector $\vX = \mM \hat{\bar{\vI}}_o$  the covariance $\mC_X\equiv \la\vX\vX^T\ra$ can be estimated from $N$ galaxy samples as
\begin{equation}
\hat{\mC}_X  = \frac{1}{N^2} \sum_i \mM(\beta_i\vI_{o,i} - \hat{\bar{\vI}}_o)(\beta_i\vI_{o,i} - \hat{\bar{\vI}}_o)^T\mM^T
\end{equation}
(for large $N$, $N \gg \text{dim}(\vX)$). The likelihood as a function of parameters $\theta$ and shear matrix $\mS$ can then be approximated as
\begin{equation}
-2\ln\cll(\mS, \theta) \sim [\hat{\bar{\vI}}_o- \vm_o(\mS,\theta)]^T\mM^T\hat{\mC}_X^{-1} \mM[\hat{\bar{\vI}}_o - \vm_o(\mS,\theta)]
\end{equation}
where $m_o(\mS,\theta) = \bar{P}\star m(\mS,\theta)$ is the model for the average PSF-convolved sheared circularly-symmetric galaxy. The simplest thing is to take $\mM$ to just re-pixelize, e.g. at the original pixel resolution. For simple PSFs it also loses little information to halve the number of points by taking $\mM$ to sum $I(\vx)$ and $I(-\vx)$ since this includes the shear and radial information, but ignores irrelevant dipole fluctuations. Note that $\mC_X$ includes variance due to both noise and `shape noise' due to the differences in galaxy shapes. The latter is expected to be spatially correlated even if the noise is not, but in any case the central limit theorem result straightforwardly accounts for any correlated or non-Gaussian noise in individual images.

In the limit that the model fits the stacked image exactly, and the stated assumptions are met, the fitting procedure should be unbiased. However due to noise this will not quite be the case, so there is potentially a source of noise-bias through the PSF and shear dependence of the estimated covariance.

 Note that even if the instrumental PSF is actually constant, the PSF for points in the stacked image plane varies from galaxy to galaxy due to the offset between the centres of the pixels in each image and the centre of the stacked image.  If the stacked image is pixelized at higher resolution, then there are different PSFs for each non-equivalent high-resolution pixel centre. However the averaged PSF will be the same for each high-resolution pixel. The high resolution pixels are of course strongly correlated due to the pixelization of the individual images.

\subsection{Centroid issues}

Centroid errors on the galaxy image plane are harmless (other than increasing the error bars), since they merely effect the average galaxy profile. For example we could consider  defining the centroid in terms of a random displacement from the centre of light in the unlensed galaxy, which is perfectly legitimate. The problem is that in general the centroid error will depend on the galaxy shape, and hence also shear and PSF in non-trivial way. Since the centroid of a long thin shape is hard to determine in the long direction, the centroid error is typically strongly correlated with the shape of the galaxy; if the galaxies have a net ellipticity in one direction due to the PSF, the centroids will tend to have a net dispersion aligned with the PSF. This has the effect of making a naively stacked image give results biased in the direction of the PSF. There are similar effects due to shear. When the centroid error is not negligible compared to the galaxy sizes, the centroid error must be accounted for somehow in order to get unbiased results. In general this is difficult, though an approximate correction may be sufficient.

Two simple approaches immediately present themself. We could simply attempt to model the effective centroid-error PSF and include it as part of the effective PSF on the stacked image. Or the centroid error could be modified to remove some of the sources of bias. The latter approach is likely to be more straightforward, if less optimal.

As a crude first attempt to remove the leading-order centroid bias I simply add Gaussian noise to each centroid so that the total centroid dispersion is approximately circularly symmetric.
To do this I fit a 6-parameter  Gaussian elliptical model to each observed (PSF-convolved) galaxy to determine the centroid, calculating the Hessian errors by numerical differentiation and then inverting to get an approximate centroid error matrix. Then to each estimated centroid I add a small Gaussian displacement in a direction chosen such that the total centroid error is then isotropic. If the estimate of the centroid error on each galaxy is fairly accurate, this should remove the correlation of centroid dispersion with galaxy alignment, and hence reduce the PSF bias. However the magnitude of the centroid error will still depend on the PSF-convolved sheared galaxy shape, and hence potentially lead to residual biases. Furthermore if the total centroid dispersion is accounted for by allowing the average galaxy profile to change,
the centroid error really has to be sheared like the rest of the galaxy shape; a better approach could therefore use an approximate estimate of the shear to ensure that the total centroid error on the image plane is sheared approximately correctly\footnote{Thanks to Gary Bernstein for pointing out this issue.}.

The centroid determined by Gaussian model fitting that I use seems to have about 10\% less dispersion than that obtained using adaptive moments~\citep{Bernstein:2001nz}, however in the presence of more complicated PSF (e.g. with a dipole) the position of the centroid could be biased, so a more sophisticated method may be required.
Unfortunately to get the centroid error correct in general requires modelling the shape of each galaxy correctly, which is just as hard as the original shape estimation problem. However as long as the centroid error is small compared to the size of the galaxies, an approximate correction may be sufficient. Indeed the output from a more realistic galaxy fitting code like Lensfit~\citep{Kitching:2008pd} might be a good place to start trying to improve the crude Gaussian model used here. Simulations may also be reliable enough to find a fudge parameter to relate the estimated centroid error to the true centroid error to the required accuracy.

\section{Results with simulations}
\label{sims}
\begin{figure}
\begin{center}
\psfig{figure=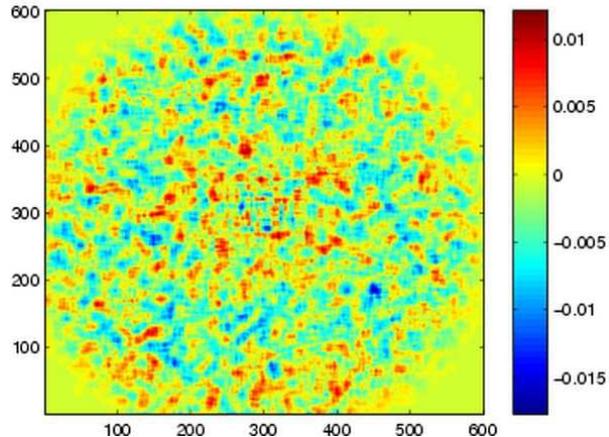,width=8cm}
\caption{Typical residuals after fitting a unit-amplitude sheared circularly-symmetric galaxy to a stacked image. Note errors are correlated due to pixel-scale stacking correlations and correlated shape noise. Here pixels have been added at 15 sub-pixel resolution, less than needed for accurate results.
\label{resid}}
\end{center}
\end{figure}

I test the galaxy stacking method on simulations provided by the GREAT08 project~\citep{Bridle:2008iv}. These satisfy the required assumptions, in that the shear is constant over a larger number of galaxy images. The simulations also have constant (very simple) PSF, with a large number of low-noise star images so that the PSF can be determined essentially exactly. The PSF is anisotropic but has no dipole moment and the isophotes have the same shape at each radius; it is therefore a rather special case and is likely to be unrealistic in several important respects. Nonetheless, even with these radical simplifications from reality, most existing shear-estimation methods fail to produce results at an accuracy required for precision cosmology, so it makes an interested test case.

The disadvantage of non-optimal methods such as stacking is there are lots of free parameters, e.g. choice of $\beta_i$ and radial fitting function, choice of the $\mM$ reduction matrix, as well as resolution parameters governing the stacking. The weights $\beta_i$ must be chosen in a shape-independent (or at least alignment-independent) manner, otherwise biases may be introduced. I take $\beta_i$ to constant or inversely proportional to the integrated signal in each images (so that the average galaxy is then independent of the magnitude distribution of the galaxies); see the Appendix for a discussion of the optimal weighting in an idealized case. For noisy images this should probably be modified by an estimate of the signal to noise ratio do down-weight noise-dominated images. I chose $\mM$ simply to re-pixelize the stacked image to the resolution of the original galaxies.

Parameterizing radial distribution of $m(\mS,\theta)$ using splines is convenient, so $\theta$ is a set of values at some radial spline nodes. Splines naturally have multiple resolution: e.g. we can do a quick fit with a few spline points, then increase the number of spline parameters to refine the result. This could be done in an adaptive way to make sure the data is fit but not over-fit.
I simply choose to spline in the log of the radial amplitude,  using 12 spline points over a radius range of 8 pixel units for fitting the stacked image, with stacked-image resolution 1/31 or  1/41 of the original pixel size (with initial fit at 1/9 resolution with 7 spline points to get close to the best-fit point quickly). The fitting could be done with MCMC to get accurate error bars,  though at a quick look does not show evidence of strong degeneracies or asymmetries in the error bars. So finding just the best fit is a reasonable first step, with errors approximated from a Hessian if required. To find the best-fit point I use the  NEWUOA\footnote{\url{http://www.inrialpes.fr/bipop/people/guilbert/newuoa/newuoa.html}} algorithm, which can be significantly faster than `AMOEBA' downhill-simplex method~\citep{Nelder65} in many cases, though I need to be a bit careful to avoid local minima. The resulting reduced chi-squared is generally less than one, indicating that indeed the galaxy is being fit accurately, though values are hard to assess due to the non-realistic mirroring procedure used in the GREAT08 simulations to help reduce shape noise. Typical residuals are show in Fig.~\ref{resid}.

For noisy images the centroid error needs to be corrected as discussed in the previous section. Comparison with the test simulation indicates that the centroid variance is underestimated by about a factor of a half, so I adopt a centroid-error fudge parameters $\alpha =1.5$ (chosen to work with the test simulations), and assume that the actual centroid covariance on each galaxy is $\alpha \mC$ where $\mC$ is estimated from the Hessian about the best-fit Gaussian model.

Accuracy of results for the purpose of GREAT08 is defined by a quality parameter $Q$~\citep{Bridle:2008iv}, so that the shear variance is
\begin{equation}
\la (\la\hat{g}_1\ra - g_1)^2+ (\la\hat{g}_2\ra - g_2)^2\ra = \frac{10^{-4}}{Q},
\end{equation}
where $\hat{g}_i$ is the shear estimated from a plate of 10000 galaxy images at constant shear and the same PSF, $g_i$ is the true shear, and $\la \hat{g}_i\ra $ is estimated from an ensemble of different simulated plates with the same shear. For noisy simulations results are quoted for $Q$ estimated from the expectation value from a set of simulations with different PSF and true shears. The target for future observations is $Q\sim {\clo (1000)}$~\citep{Amara:2007as}, and current methods typically give $Q \alt \clo(100)$. Biases on $g_i$ therefore need to be below $\alt 3\times 10^{-4}$ level, or a typical fractional shear error of less than about a percent. For low-noise images the definition is simply to take each plate separately
\begin{equation}
\hat{Q} = \frac{10^{-4}}{\la (\hat{g}_1 - g_1)^2+ (\hat{g}_2 - g_2)^2\ra_{\text{plates}} },
\end{equation}
with the stacking method described here giving $\hat{Q} \sim 300$.  The errors have a contribution from any systematics error and intrinsic shape noise (which may be significantly higher than possible due to the lossy nature of the stacking procedure).  Most methods used with current data give $Q \alt 30$.

When the noise is significant the method is no longer strictly valid due to centroid issues, however using the centroid error correction described above still gives $\hat{Q} \sim 130$, which is at as good as other existing methods at the time of this work, and within a factor of two of the best method eventually winning the GREAT08 challenge. However the stacking method is more reliant on the non-realistic constant-shear assumption than some other methods, so the main use may be as a baseline for simulation-based comparisons with better codes.

The fact that $Q\agt 100$ can be obtained by this sub-optimal method, making essentially no assumptions about the galaxy distribution, is perhaps encouraging evidence that there will exist a better method that is good enough for precision cosmology using only modest assumptions about the galaxy distribution.
There is some evidence for shear-calibration bias in the stacking results, with a tendency for $|g|$ to be too large. More careful modelling of the centroid error, for example using better model fitting and an iterative shear estimate, could probably reduce the systematic error. Suitable time-consuming adjustment of the method parameters may also allow the method to perform significantly better.

\section{Conclusions}

I have revisited the simple shear-estimation stacking method of~\citet{Kuijken:1999jz}, and shown that it still makes a useful baseline that can compare favorably with currently used methods in idealized cases. Although it only works straightforwardly over regions with constant shear, it can be a useful test case, and help to understand possible sources of bias in other methods.  Stacking has the advantage of giving results that are unbiased almost independent of the unknown distribution of unlensed galaxy shapes. Residual biases enter at a lower level, for example through correlations of the centroid error with galaxy shape. With low noise the method can produce accurate results, comparing favourably with methods that fit galaxy models that are known to be unrealistic. This should be unsurprising: Bayesian methods generally give the right results only if the correct model is used and priors truly reflect beliefs.  Only in the very special case of no observational PSF does fitting any model to galaxy shapes give unbiased answers; a general PSF breaks all the symmetries, requiring accurate modelling of both the PSF and the unlensed galaxy shape distribution to get the right result.  

Even if individual noisy galaxies are well fit by a simple galaxy model due to the large noise, if in reality galaxies have significant substructure or un-modelled shape variations, the combined high-precision shear estimate from fitting many galaxies separately may be biased due to the inconsistent shape modelling. It is possible the modelling bias is negligible, but unless carefully proven analytically or demonstrated numerically in realistic simulations it would be safer to assume otherwise (see~\citet{galaxies2009} for a quantitative analysis of the significant bias in various idealized cases). The noise-free stacking procedure is by construction linear in the galaxies, which is why substructure variations between galaxies effectively cancel. However fitting to individual galaxies is usually a non-linear procedure, and there is no reason to expect errors to cancel more generally. Future work may however be able to find fairly model-independent methods that can be applied to fitting individual galaxies, significantly improving on the stacking method both in terms of signal to noise, and in terms of application to more realistic cases. If not, stacking methods may still be useful. Future work could investigate how to apply stacking in more realistic cases where the shear varies from galaxy to galaxy, and the PSF can only be estimated locally with significant noise. At leading order, a fit to a stacked galaxy constructed over a region with small variations in shear should be probing the appropriately averaged shear. With a high galaxy density the corresponding suppression of small-scale power may be acceptable if it can be accurately modelled without significant bias.

\section{Acknowledgements}
I acknowledge a PPARC/STFC Advanced fellowship and thank members of the GREAT08 team, especially Sarah Bridle, for helpful discussions, Lance Miller for comments, Steven Gratton for matrix identity advice, and CITA for use of their old computer.

\appendix

\section{Analytic example}

Here we consider a very simple toy distribution of galaxies where we can attempt to calculate some things analytically. Consider the case where each galaxy has a Gaussian-shaped profile
\begin{equation}
I_{u}(\vx|\mQ_i) = A \frac{e^{-\vx^T\mQ_i^{-1}\vx/2}}{|\mQ_i|^{w/2}},
\end{equation}
where the distribution of the covariance $\mQ_i$ of each galaxy is drawn from a 2-dimensional inverse Wishart distribution (see e.g.~\citet{Gupta99} for review and results used below)
\begin{equation}
P(\mQ) = \frac{|\mpsi|^{(n-3)/2} e^{-\half \Tr(\mpsi \mQ^{-1})}}{2^{n -3}\,
\pi^{1/2}\Gamma[(n-3)/2]\Gamma[(n-4)/2]
|\mQ|^{n/2}}
\end{equation}
where $n > 6$.
  Since we assume the unlensed distribution is statistically isotropic $\mpsi_0= (n-6)\sigma_g^2 \mI$ where $\sigma_g$ is the average galaxy width. The parameter $n$ determines how broad the galaxy shape distribution is, with $n\rightarrow \infty$ corresponding to a distribution of identical circular Gaussian galaxies.
  Typical galaxy ellipticities are $\clo(n^{-1/2})$ with
\begin{equation}
\frac{\la (Q_{11}-Q_{22})^2\ra}{\la (Q_{11}+Q_{22})^2\ra} =
\frac{\la 4Q_{12}^2\ra}{\la (Q_{11}+Q_{22})^2\ra} =
 \frac{1}{n-6}.
\end{equation}
The parameter $w$ governs how the magnitude varies, with $w=0$ corresponding to all galaxies having the same peak amplitude, and $w=1$ corresponds to them all having equal integrated light.

The averaged galaxy profile (with equal weight) is given by
\begin{eqnarray}
\bar{I}_u(\vx) &=&  \int \ud \mQ P(\mQ) I_u(\vx|\mQ)\nonumber \\
&=& \!\!\frac{\Gamma[n+w-4]}{\Gamma[n-4]|\mpsi|^{w/2}}\frac{A}{ (1 + \vx^T \mpsi^{-1}\vx)^{(n+w-3)/2} }.
\end{eqnarray}
As expected this becomes the same as the individual galaxy shape as $n\rightarrow \infty$.
The covariance can be calculated similarly as
\begin{multline}
\text{cov}(\vx,\vy) =  \int \ud \mQ P(\mQ) I_u(\vx|\mQ)I_u(\vy|\mQ) -
\bar{I}_u(\vx)\bar{I}_u(\vy) \\
= \frac{A^2}{\Gamma[n-4]|\mpsi|^w}\bigg[ \frac{\Gamma[n+2w-4]}{|\mI+ \mpsi^{-1}(\vx\vx^T +\vy\vy^T)|^{(n+2w-3)/2}}
\\
 - \frac{\Gamma[n+w-4]^2}{\Gamma[n-4]  [(1 + \vx^T \mpsi^{-1}\vx)(1 + \vy^T \mpsi^{-1}\vy)]^{(n+w-3)/2} }\biggr].
\label{analytic_cov}
\end{multline}
Note that
\begin{multline}
|\mI+ \mpsi^{-1}(\vx\vx^T +\vy\vy^T)| =\\(1 + \vx^T \mpsi^{-1}\vx)(1 + \vy^T \mpsi^{-1}\vy) - (\vx^T\mpsi^{-1}\vy)^2.
\end{multline}
The covariance is determined by the number of degrees of freedom governing the population, so that with a simple model the number of significantly non-zero eigenvalues is small. In the case here each galaxy shape is determined by the three independent numbers in $\mQ$.

Using this analytic galaxy population model we can compare the errors (e.g. estimating the shear matrix $\mS$ such that $\mpsi = \mS^T\mpsi_0\mS$)  using stacking compared to what could be done using an optimal analysis. If $\mQ_i$ were simply measured directly from each galaxy (in the low-noise limit) then the optimal expected error is
\begin{equation}
\left\la \sum_i \frac{\partial^2 \ln P(\mQ_i)}{\partial g_a\partial g_{b}}\right\ra^{-1}_{g=0} = \frac{\delta_{ab}}{4N(n-3)},
\end{equation}
where $N$ is the number of galaxies. The corresponding error per galaxy is  $\sigma_{1} = \sigma_{2}= 1/(2\sqrt{n-3})$, where $\sigma_i$ is the error on $g_i$. For $n=7$ this corresponds to an error per galaxy $\sigma_i = 0.25$.

Using stacked galaxies with $w=0$ in fact gives the same average error per galaxy, with the errors increasing only slightly for $w\sim\clo(1)$. In this noise-free case with known distributions and centroids, the stacking method is close to optimal. To show that with $w=0$ stacking gives optimal answers we only need to show that the full likelihood can be written in terms of the stacked image. Since
\begin{equation}
-2  \ln P(\mpsi|\{\mQ_i\}) = \sum_i \left[\Tr(\mpsi\mQ_i^{-1}) - (n-3)\ln|\mpsi|\right] + \text{const},
\end{equation}
where the last term is independent of $\mpsi$, a sufficient statistic is $\sum_i \mQ^{-1}_i$.  However this can be measured by taking derivatives of the perfect stacked image
\begin{equation}
\hat{\bar{I}}_u = \frac{A}{N}\sum_i e^{-\vx^T\mQ_i^{-1}\vx}
\end{equation}
at the origin, and hence stacking is lossless for measuring the shear in this ideal case. Since the number of degrees of freedom in the galaxy model is small, the stacked image does not actually need to be densely sampled to obtain close to optimal results.

In the zero-noise limit with infinite resolution, a known PSF can simply be deconvolved, so the above results also apply to PSF-smeared noise-free galaxies. Noise can be accounted for by adding an appropriate term to Eq.~\eqref{analytic_cov} if the centroids are known, and will increase the expected error per galaxy. Analysing more realistic cases analytically is challenging.


\bibliography{../antony,../cosmomc}
\bibliographystyle{mn2e_eprint}

\end{document}